# Magnetospherically Driven Optical and Radio Aurorae at the End of the Main Sequence

G. Hallinan, S. P. Littlefair, G. Cotter, S. Bourke, L. K. Harding, J. S. Pineda, R. P. Butler, A. Golden, G. Basri, J.G. Doyle, M. M. Kao, S. V. Berdyugina, A. Kuznetsov, M. P. Rupen & A. Antonova

**Affiliations**

G. Hallinan, S. Bourke, J. S. Pineda & M.M. Kao

*California Institute of Technology, 1200 East California Boulevard, Pasadena, California 91125, USA*

S. P. Littlefair

*Department of Physics and Astronomy, University of Sheffield, Sheffield S3 7RH, UK.*

G. Cotter

*University of Oxford, Department of Astrophysics, Denys Wilkinson Building, Keble Road, Oxford OX1 3RH*

L. K. Harding

*Jet Propulsion Laboratory, California Institute of Technology, 4800 Oak Grove Drive, Pasadena, CA 91109-0899, USA*

R. P. Butler

*Centre for Astronomy, National University of Ireland-Galway, University Road, Galway, Ireland*

A.Golden

*Department of Mathematical Sciences, Yeshiva University, New York, NY 10033, USA*

G. Basri

*Astronomy Department, University of California, Campbell Hall, Berkeley, CA 94720, USA*

J. G. Doyle

*Armagh Observatory, College Hill, Armagh BT61 9DG, N. Ireland*

S. V. Berdyugina

*Kiepenheuer Institut für Sonnenphysik, Schöneckstrasse 6, D-79104 Freiburg, Germany*

A. Kuznetsov

*Institute of Solar-Terrestrial Physics, Irkutsk664033, Russia*

M. P. Rupen

*National Radio Astronomy Observatory, P.O. Box O, Socorro, NM 87801*

A. Antonova

*Department of Astronomy, Faculty of Physics, St Kliment Ohridski, University of Sofia, 5 James Bourchier Boulevard, 1164 Sofia, Bulgaria*

**Aurorae are detected from all the magnetized planets in our Solar System, including Earth[1]. They are powered by magnetospheric current systems that lead to the precipitation of energetic electrons into the high-latitude regions of the upper atmosphere. In the case of the gas giant planets, these aurorae include highly polarized radio emission at kHz and MHz frequencies produced by the precipitating electrons[2], and a myriad of continuum and line emission in the infrared, optical, ultraviolet and X-rays associated with the collisional excitation and heating of the hydrogen-dominated atmosphere[3]. Here we present simultaneous radio and optical spectroscopic observations of an object at the end of the main sequence, located right at the boundary between stars and brown dwarfs, from which we have detected radio and optical auroral emissions both powered by magnetospheric currents. Whereas the magnetic activity of stars like our Sun is powered by processes that occur in their lower atmospheres, these aurorae are powered by processes originating much further out in the magnetosphere of the dwarf that couple energy into the lower atmosphere. The dissipated power is at least $10^4$ times larger than produced in the Jovian magnetosphere, revealing aurorae to be a potentially ubiquitous signature of large-scale magnetospheres that can scale to luminosities far greater than observed in our Solar System. These magnetospheric current systems may also play a causal role in some of the reported weather phenomena on brown dwarfs.**

LSR J1835+3259 (hereafter LSR J1835) is a dwarf of spectral type M8.5 with a bolometric luminosity $10^{-3.4}$ times that of the Sun, located at a distance of 5.67 +/- 0.02 pc[4]. It is positioned close to a transition in magnetic activity near the end of the main sequence, where the fractional X-ray luminosity ($L_x/L_{bol}$), indicative of the presence of a magnetically heated corona, drops by two orders of magnitude over a small range in spectral type[5]. Simultaneously, rapid rotation becomes ubiquitous, indicating a dearth of stellar wind assisted magnetic braking[6]. Together, these results suggest that the coolest stars and brown dwarfs possess a comparatively cool and neutral outer atmosphere relative to earlier type dwarf stars. Consistent with this picture, LSR J1835 is a rapid rotator with a period of rotation of just 2.84 hours, and previous deep *Chandra* observations have failed to detect any X-ray emission associated with the presence of a magnetically heated corona[7].

LSR J1835 has previously been identified to produce highly circularly polarized radio emission, periodically pulsed on the rotation period of 2.84 hours[8]. Since their initial detection as a new population of radio sources[9], similar behaviour has been observed for a number of very low mass stars and brown dwarfs spanning the spectral range M8 - T6.5[10,11]. In some cases, periodic variability has also been detected in broadband optical photometric bands and the Hα line[12,13,14]. Together these characteristics are unlike anything observed for earlier type stars[15].

We pursued spectroscopic data in radio and optical bands to investigate a possible relationship between the periodic radio, broadband optical and Balmer line emission. We used the Karl G. Jansky Very Large Array (VLA) radio telescope to produce a dynamic spectrum of the periodic radio emission from LSR J1835. Simultaneously we conducted time resolved optical spectrophotometry using Double Spectrograph (DBSP) on the 5.1m Hale telescope at the Palomar Observatory. Follow-up observations, involving an additional 7 hours of more sensitive time resolved optical spectrophotometry, were carried out using the Low Resolution Imaging Spectrometer (LRIS) on the 10 m Keck telescope.

The broadband ($\delta v/v \sim 1$) dynamic radio spectrum produced with the VLA reveals a number of distinct components periodically repeating on the 2.84 hour rotation period (Figure 1). The observed periodic features can exhibit a cut-off in frequency, are 100% circularly polarized and are very short duration relative to the rotation period, the latter implying sharp beaming. These properties are consistent with electron cyclotron maser emission produced near the electron cyclotron frequency at the source of the radio emission ($v_{MHz} \approx 2.8 \times B_{Gauss}$), a coherent emission process responsible for planetary auroral radio emission[2]. From the dynamic spectrum of the radio emission from LSR J1835, we can infer magnetic field strengths in the source region of the emission ranging from 1550 to at least 2850 Gauss, close to the maximum photospheric magnetic field strengths found in late M dwarfs[16]. The proximity of the radio source close to the photosphere, together with the persistent nature of the periodic radio emission, requires a current system driving a continuously propagating electron beam in the lower atmosphere of the dwarf.

The simultaneous optical spectroscopic data collected with the Hale telescope are also modulated on the 2.84 hour rotational period (Fig. 1). This behaviour is confirmed with the follow-up higher signal-to-noise Keck data, where the same periodic modulation is observed at the same amplitude. This modulation is clearly present in both spectral line emission, including the Balmer lines, and the broadband continuum optical emission of the dwarf. Most notably, the Balmer line emission and nearby continuum vary in phase (Fig. 2), indicating a co-located region of origin.

We find that the surface feature responsible for the periodic variability in the optical spectrum can be modelled as a single component approximated as a blackbody of temperature T ~ 2200 K, with surface coverage of < 1% (Fig. 2). We attribute this blackbody-like spectrum to an optically thick region with the dominant opacity contributed by the negative hydrogen ion (H-). H- is the dominant source of solar optical continuum opacity, but is superseded by bound-bound molecular line opacities, such as TiO, for cool M dwarfs due to the scarcity of free electrons available to form the H- ion[17]. In the solar case, ionized metals fulfil the role of the electron donors necessary to sustain a H- population. Since there are essentially no electron donors in a thermal gas at T = 2200 K, another population of electron donors is required.

The periodically variable radio, Balmer line and optical continuum emission detected from LSR J1835 can be explained by a single phenomenon, specifically a propagating electron beam impacting the atmosphere, powered by auroral currents. Integrating over time and frequency, we can determine that the highly circularly polarized radio emission from LSR J1835 contributes at least $10^{15}$ W of power, requiring $10^{17} - 10^{19}$ W of power available in the electron beam for dissipation in the atmosphere, assuming an efficiency of $10^{-2} - 10^{-4}$ for the radio emission[2]. We note that this amounts to ~$10^{-6} - 10^{-4}$ of the bolometric luminosity of the dwarf. Collisional excitation of the neutral hydrogen atmosphere by the precipitating electrons leads to subsequent radiative de-excitation, resulting in Balmer line emission with an average power of 2.5 x $10^{17}$ W, consistent with the energy budget of the precipitating electron beam. This is similar to the main Jovian auroral oval, where, radio emission from electron beams contributes only $10^{-4}$ of the auroral power, with the bulk of the power produced in the infrared ($H3+$; thermal), far ultraviolet (Lyman and Werner band $H_2$ emission) and optical (Balmer line) due to dissipation of the electron beam energy in the atmosphere[3]. Similar ratios are also observed in Io's magnetic footprint in the Jovian atmosphere[18].

We propose that this same electron beam is also causally responsible for the co-located optical broadband variability. In this model the associated increased electron number density contributes excess free electrons leading to an increase in the H- population, in a process that has previously been invoked for white light solar flares[19]. This results in an optically thick layer at a higher altitude, and thus lower temperature, than the photosphere. Despite the lower temperature, the absence of deep absorption bands in the spectrum results in a bright feature in optical bands, responsible for the broadband optical variability. However, in regions of the dwarf spectrum devoid of absorption bands, particularly towards the redder end of the spectrum, there is a reversal, with the auroral feature appearing dimmer than the surrounding photosphere. This results in some optical wavelengths displaying lightcurves which are in anti-phase to other wavelengths (Fig 2). This phenomenon also accounts for the anti-phased lighturves seen in multi-band photometry of an M9 dwarf[13].

Our observations point to a unified model involving global auroral current systems to explain the periodic radio, broadband optical and Balmer line emission detected from LSR J1835, as well as other low mass stars and brown dwarfs. Extending to cooler objects, it is notable that radio emission has now been detected from a brown dwarf of spectral class T6.5 (~900 K)[11]. The brown dwarf in question is also notable for hosting weak Hα emission, one of only 3 such T dwarfs confirmed to emit Balmer line emission[20]. The similarities with LSR J1835 suggest that the auroral mechanism robustly operates well into the regime occupied by the coolest brown dwarfs of spectral types late L and T. It is also notable that a large degree of variability has been observed in the infrared in this spectral regime, particularly near the transition between the L and T spectral classes[21,22]. This variability explicitly requires variation in temperature or photospheric opacity across the surface of these brown dwarfs[23], which has been attributed to the spatially inhomogeneous distribution of condensate clouds in their atmospheres, effectively a manifestation of weather. This is supported by the mapping of such cloud patterns on the surface of the recently discovered Luhman 16B[24]. We speculate that the magnetospheric currents powering aurorae in brown dwarfs may also play a causal role in driving some of the more extreme examples of the weather phenomenon in brown dwarfs; specifically, via downward propagating electron beams modifying atmospheric temperature and opacity in the same fashion as has been shown for LSR J1835.

The nature of the electrodynamic engine powering brown dwarf aurorae remains an outstanding open question. For solar system planets, this electrodynamic engine can be a) magnetic reconnection between the planetary magnetic field and the magnetic field carried by the solar wind (e.g. Earth and Saturn)[25] b) the departure from co-rotation with a plasma sheet residing in the planetary magnetosphere (e.g. Jovian main auroral oval)[26,27] or c) interaction between the planetary magnetic field and orbiting moons (e.g. Jupiter-Io current system)[28]. Of these models, the sub-corotation of magnetospheric plasma on closed field lines, in turn powering magnetosphere–ionosphere coupling currents, has been put forward as a plausible model that can be extrapolated from the Jovian case to the brown dwarf regime[29, 30]. This latter model requires a continuously replenished body of plasma within the magnetosphere. This mass-loading can be achieved by multiple avenues, including interaction with the ISM, the sputtering of the dwarf atmosphere by auroral currents, a volcanically active orbiting planet or magnetic reconnection at the photosphere. Alternatively, considering the case of an orbiting planetary body embedded within the magnetosphere of LSR J1835, a simple scaling from the Jupiter-Io system indicates that an Earth-sized planet (magnetized or unmagnetized) orbiting within 20 radii (<30 hour orbital period) of LSR J1835 will generate a current sufficient to power the observed aurorae. However, we note that the observed rotational modulation of the radio emission would require a substantially asymmetric magnetic field configuration for LSR J1835. Indeed, the aurorae would display modulation on both the rotational and orbital period, which may be consistent with the large degree of variability reported for the radio pulsed emission from these objects.

A possible avenue to resolving the nature of the electrodynamic engine lies with the strong rotational modulation of the Balmer line emission of LSR J1835, which implies the auroral feature is not axisymmetric relative to the rotation axis of the dwarf. This should result in a variation in the width, intensity and velocity structure of line profiles with rotation that can be used to help map the aurora (e.g. auroral oval vs polar cap), analogous to Doppler imaging, which in turn will inform on the location of the electrodynamic engine.

Our results imply that the available power for generating aurorae on brown dwarfs is dependent on magnetic dipole moment and rotation, and may be weakly coupled to other physical characteristics, such as bolometric luminosity. This accounts for the continuous presence of auroral radio emission at similar luminosity from spectral type M8 through to T6.5, despite a 2 orders of magnitude decrease in bolometric luminosity over the same spectral range[8]. This suggests aurorae may be present at detectable levels on even the faintest T and Y dwarfs and bodes well for searches for similar emission from exoplanets.

**Online Content** Methods, along with any additional Extended Data display items and Source Data, are available in the online version of the paper; references unique to these sections appear only in the online paper.

**Supplementary Information** is linked to the online version of the paper at www.nature.com/nature.

**Acknowledgements** We are grateful to T. Readhead, S. Kulkarni and J. McMullin for working to ensure that simultaneous Palomar and VLA observations could occur. We thank the staff of the Palomar Observatory, the W.M. Keck Observatory and the National Radio Astronomy Observatory for their support of this project. The W.M. Keck Observatory is operated as a scientific partnership among the California Institute of Technology, the University of California and the National Aeronautics and Space Administration. The Observatory was made possible by the generous financial support of the W.M. Keck Foundation. The VLA is operated by the National Radio Astronomy Observatory, a facility of the National Science Foundation operated under cooperative agreement by Associated Universities, Inc. Armagh Observatory is grant-aided by the N. Ireland Dept. of Culture, Arts & Leisure. G.H. acknowledges the generous support of D. Castleman and H. Rosen. This material is based upon work supported by the National Science Foundation under grant no. AST-1212226/DGE-1144469. We thank Jeff Linsky and Peter Goldreich for helpful comments on this manuscript.

**Contributions**

G.H., S.B., M.R., A.A., A.G., A.K., M.M.K. and J.G.D. proposed, planned and conducted the radio observations. G.H. and S.B. reduced the VLA data and the dynamic spectrum was outputted by S.B. G.H. interpreted the dynamic radio spectra. G.H., S.P.L., G.C., R.P.B., S.P. and L.K.H. proposed and conducted the Keck observations. G.C. carried out the Palomar observations and reduced the publication data. S.P.L. and G.C. reduced the Keck spectroscopic data, with the final publication data delivered by S.P.L. G.H. G.C. and S.B. G.H., S.P.L. and J.S.P developed the interpretation of the optical data. S.P.L. carried out the detailed model fitting of the Keck spectra. G.B. analyzed high resolution archival spectra and provided insight on interpretation of the optical data. S.V.B. coordinated contemporaneous spectropolarimetry with the observations presented in this paper. S.P.L and G.H. wrote the Supplementary Information. All authors discussed the result and commented on the manuscript.

**Author Information**

Reprints and permissions information is available at www.nature.com/reprints. The authors declare no competing financial interests. Correspondence and requests for materials should be addressed to G.H. (gh@astro.caltech.edu).

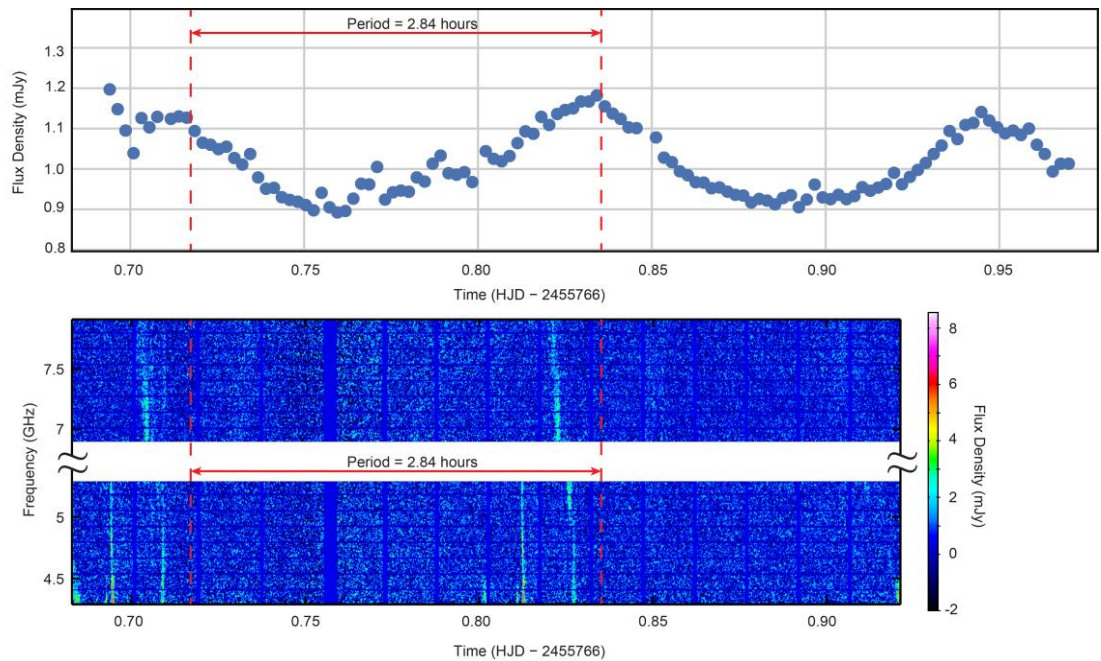

**Figure 1 | Simultaneous optical and radio periodic variability of LSR J1835. a.** Balmer line emission extracted from spectra detected with the Hale telescope. **b.** Dynamic spectra of the right circularly polarized radio emission detected from LSR J1835 with the VLA, with the y axis truncated to remove the large gap between observing bands (see Methods for details). The offset in phase of the radio features relative to the optical peak can be accounted for by the complex beaming of the radio emission.

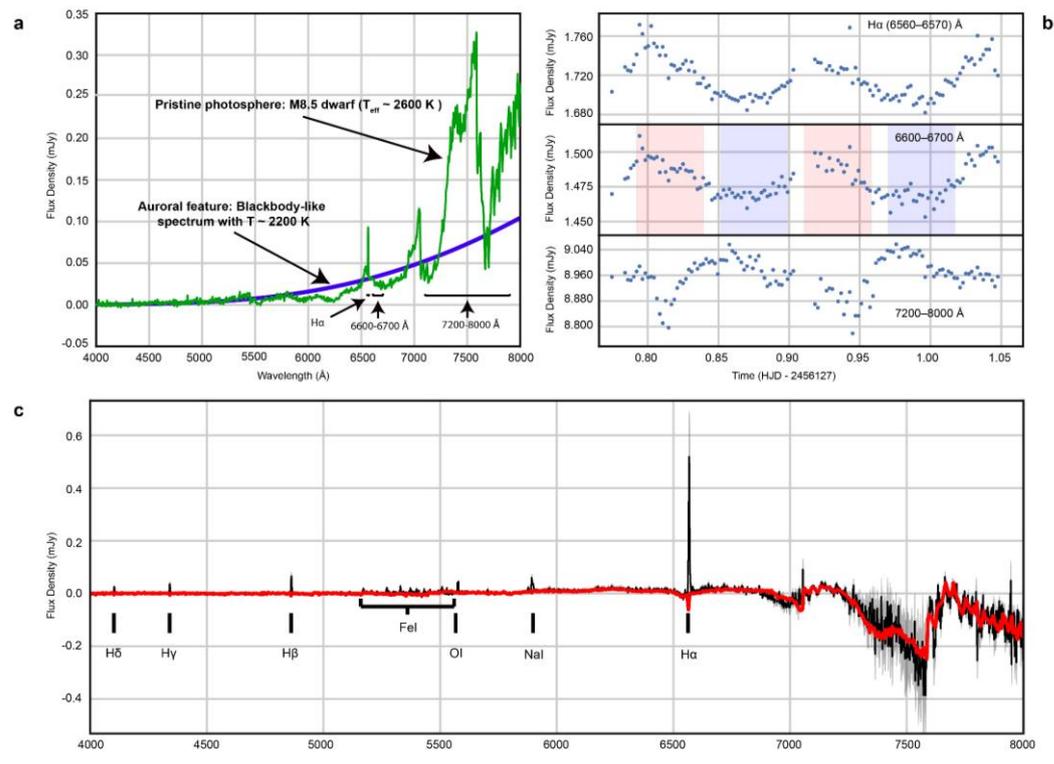

**Figure 2 | Modelling the optical variability of LSR J1835. a.** Adopted models for the surface brightness of the pristine photosphere (green) and auroral feature (blue). At certain wavelengths the auroral feature is brighter, whereas the pristine photosphere is brighter at other wavelengths. **b.** Lightcurves constructed from the Keck spectrophotometry for three regions of the spectrum highlighted in panel **a**. The lightcurve produced for Hα emission is tightly correlated with the lightcurve of the nearby continuum, confirming the Balmer line emission and excess continuum emission to be approximately co-located. Meanwhile, redder wavelengths are in anti-correlation as expected for our model. **c.** The amplitude of variability as a function of wavelength (black) with 2σ uncertainties (shaded gray region), derived from the 'low' and 'high' states for the auroral emission defined as the red and blue shaded regions respectively in panel **b**. The variability predicted by the model presented in panel **a**, is shown in red (see Methods for details). We note that the line emission is not represented in the model.

**Methods**

**Radio Data Reduction**

Data were reduced using the Common Astronomy Software Applications (CASA Release 4.1.0) and Astronomical Image Processing System (AIPS) packages. The amplitude and phase of the data were calibrated using short observations of the quasars QSO J1850+284 and 3C286 that were interspersed throughout the 6 hour observation of LSR J1835. Bad data, particularly those contaminated by radio frequency interference (RFI), were flagged. The tasks *fixvis/UVFIX* were used to shift the source to the phase center and the tasks *clean/IMAGR* were used to image the data. The tasks *uvsub/UVSUB* were used to subtract the source models for nearby background sources from the visibility data. The real part of the complex visibilities as a function of time and frequency for each polarization were then exported from CASA and AIPS and plotted to produce the dynamic spectrum of LSR J1835 shown in the main paper.

We observed LSR J1835 using 2 x 1 GHz sub-bands spanning frequencies of 4.3-5.3 GHz and 6.9-7.9 GHz. Two full rotation periods were captured during the 5.7-hour observation. We show the dynamic spectrum detected by the right circularly polarised feeds of the VLA antennas. The original data has a time and frequency resolution of 1 second and 2 MHz respectively, but is binned and smoothed to produce the data shown in Figure 1, with the y axis truncated to remove the large gap between observing bands. Periodic features of 100% circularly polarized radio emission occupy ~3% of the dynamic spectrum, allowing us to infer that it is beamed from the source in an angular emission pattern that occupies a similar fraction of a 4π steradian sphere. Studies of electron cyclotron maser emission from planetary magnetospheres have revealed the emission to be beamed in a hollow cone with walls a few degrees thick and a large opening angle (typically ~ 70° to 90°) relative to the local magnetic field[2]. This is consistent with our data and accounts for the offset in phase of the radio features relative to the optical peak. Integrating over time and frequency and inferred beaming pattern, we determine the auroral power contributed by the polarized radio emission to be ~$10^{15}$ W.

We detect at least 6 distinct components in the dynamic spectrum for LSR J1835, each of which is likely powered by a distinct local field-aligned current. The relationship between these individual current systems within the large scale magnetosphere will likely remain uncertain until the nature of the electrodynamic engine is established. Although a number of these components exhibit a cut-off in emission rising to higher frequencies, there are still components present all the way to the top of the band, implying that the true cut-off in emission, associated with the largest strength magnetic fields near the surface of the dwarf, was not captured.

**Hale Optical Data Reduction**

The data from the Double Spectrograph (DBSP) on the Hale telescope were reduced using standard techniques with the aid of the IRAF software suite. First, bias level was subtracted from the raw frames. Then pixel-to-pixel gain variations in the CCD were corrected by normalising against exposures taken with the DBSP internal broadband lamp and the illumination function of the long slit was corrected by normalising against twilight sky exposures. Next the {x,y} pixels of the CCD were transformed to a rectilinear {wavelength, sky position} solution using the DBSP internal arc lamps.

Finally the night sky emission was subtracted by fitting a fourth-order polynomial to each column along the sky direction, with the stars in the frame masked out, and cosmic rays rejected via sigma-clipping. A tramline extraction was then used to make 1-D spectra of the target and reference stars.

**Keck Optical Data Reduction**

Data were reduced using the LOW-REDUX pipeline (http://www.ucolick.org/~xavier/LowRedux). Bias frames were constructed by median stacking 5 individual bias frames. Non-uniform pixel response was removed by dividing by a dome flat field produced by stacking 7 individual flat fields together. Individual objects were located on the slit, and an optimal extraction routine[31] was used to extract the object spectra on each frame. Wavelength calibration was performed using fits to arc line spectra taken at the start of the night, which gave a dispersion of 2.4 AA/pix (red) and 3.2 AA/pix (blue), and RMS values of 0.7 pix (red) and 0.4 pix (blue). Each object spectrum was corrected for flexure using fits to night sky lines. Flexure corrections ranged from -2 to +2 pixels. Flux calibration was performed via a high-order polynomial fit to the flux standard Feige-110, with the Balmer lines masked out.

The data were corrected for light falling outside of the spectrograph slit using the additional comparison stars observed. Since the wavelength coverage of each object differs slightly due to the location of slits in the object mask, slit loss corrections were determined as follows. Each comparison star was divided by an average of all the frames, to remove the spectral shape. The resulting spectrum was fit with a 1$^{st}$ order polynomial to give a series of wavelength-dependent slit loss corrections for each frame. This polynomial was then re-binned onto the same wavelength scale as the target spectrum. Not all comparison star spectra were used to correct for slit losses in the target spectrum. Instead, a master slit loss correction was produced via a straight mean of slit loss corrections for selected comparison stars, with the quality of slit loss correction being judged by-eye. Comparison stars were removed from the slit loss correction calculations because they were either extremely blue, or not well aligned on their slits in the slit mask. LSR 1835 was slit loss corrected using the two reddest comparison stars.

## Wavelength dependence of optical variability

The amplitude of variability as a function of wavelength (which we term the difference spectrum) was estimated as follows.

A rotational phase was assigned to each spectrum using the ephemeris of LSR J1835. We created a spectrum representing the "high state" and the "low state" of LSR J1835 by averaging all spectra with phases between 0.95-1.00 & 0.00-0.35 and 0.45-0.85 respectively. The difference spectrum is then simply the difference between the high state and the low state spectra.

Statistical uncertainties on the difference spectrum are negligible compared to systematic errors which arise from imperfect correction of slit losses, sky subtraction and removal of telluric absorption. To estimate these systematic errors, we produced a difference spectrum using an independent method. In this method, lightcurves were produced from a series of 5Å bins, and a sinusoid of fixed phase and period was fitted to the lightcurves. A difference spectrum was produced using the amplitude of the sinusoid fit at each wavelength. The two methods yield very similar spectra, except in the range between 7600 and 7650 AA, where the spectra are affected by telluric absorption. Subtracting the two difference spectra gave an estimate of the uncertainty, which is shown in Fig. 2.

**Modelling the variability**

We construct a two-phase model of the optical emission from LSR J1835, with emission from a 'pristine' photosphere P, and a surface feature S. We assume that the relative contribution from these two phases varies as the dwarf rotates. If the surface feature covers a fraction $f_h$ during the high state and a fraction $f_l$ during the low state, then the difference between high and low states can be written as

$$\Delta = f_h S + (1 - f_h)P - f_l S - (1 - f_l)P,$$

which can be simplified to

$$\Delta = (f_h - f_l)(S - P) = \epsilon(S - P).$$

We model the photosphere, P, of LSR J1835 using an M8 template from a SDSS library of composite M-dwarf spectra[32]. To ensure the absolute surface flux of the template is correct, we scale the template so that the bolometric flux matches that of a DUSTY model atmosphere[33] with surface gravity of log g=5.0 and an effective temperature 2600K. We model the surface feature, S, as a black body with temperature $T_b$. Since our models for S and P give the flux crossing unit surface of the star, to match them to our data they need to be multiplied by a factor N = (R/d)$^2$, where R is the radius of LSR J1835 and d is the distance to LSR J1835. Since this is a simple multiplication of the model, this factor can be combined with the parameter $\epsilon$. The two free parameters of our model are therefore Tb, and the scaling constant, $\epsilon$.

We draw samples from the posterior distributions of our parameters by a Markov-Chain Monte Carlo (MCMC) procedure. Because our difference spectrum has unknown uncertainty, a nuisance parameter σ is added. The uncertainty on the difference spectrum is set to σ everywhere, except at wavelengths corresponding to emission lines where the uncertainty is set to an arbitrary large value.

Posterior probability distributions of Tb, $\epsilon$ and σ are estimated using an affine-invariant ensemble sampler[34]. Uninformative priors were used for all parameters, with the exception that $\epsilon$ was forced to be positive. The MCMC chains consist of a total of 48,000 steps of which 24,000 were discarded as burn-in, giving 560, 770 and 860 independent samples of Tb, $\epsilon$ and σ, respectively. The posterior probability distributions of our parameters are shown in Extended Data Fig. 1. Chisq of the most probable model was 860, with 920 degrees of freedom, showing that the model is an excellent fit to the data. The only wavelength regions where the model fails to reproduce our data is in the emission lines. The emission lines are likely caused by collisional excitation of the neutral atmosphere by the precipitating electrons leads to subsequent radiative de-excitation; a process not captured by our simple two-phase model.

The best fitting parameters are Tb = 2180±10K, $\epsilon = (1.64 \pm 0.02) \times 10^{-21}$ and $\sigma = 0.018 \pm 0.0004$ mJy. Using a model-based estimate for the radius and the measured distance for LSRJ1835 to correct $\epsilon$ for the (R/d)² factor, we find that the difference in covering fractions between the high state and low state is between 0.5 and 1%. These error bars do not take into account systematic uncertainties. For example, the photospheric temperature of LSR J1835 is not determined to 10K; adopting a different template, of M9 spectral type, for the pristine photosphere can alter Tb by approximately 50K. Similarly, systematic errors in the scaling factor $\epsilon$ are probably around five times higher than the statistical errors quoted above.

## Modelling the high state

By fitting the difference spectrum we are able to constrain the auroral surface feature's spectrum with a minimum of assumptions. Nevertheless, to give confidence in our modelling, one might wish to compare our observed high state spectrum with the predictions of our model. This requires a few additional assumptions. The high state can be written as

$$H = N[f_h S + (1 - f_h)P].$$

Assuming the same pristine spectrum P and surface feature spectrum S that gave the best fit to our difference spectrum, we use an identical MCMC procedure to that outlined above (including a similar nuisance parameter $\sigma$ for the uncertainties) to draw posterior samples of N and $f_h$. We find $f_h = 0.024 \pm 0.004$, $N = (8.47 \pm 0.02) \times 10^{20}$ and $\sigma = 0.196 \pm 0.003$ mJy. The resulting fit to the high state spectrum and residuals are shown in Extended Data Fig. 2. The constraints on $N$ above, and the constraint on $\epsilon$ from fitting the difference spectrum allow us to estimate $f_h - f_l = 0.0194 \pm 0.0002$ and hence $f_l = 0.005 \pm 0.004$.

A couple of pertinent features are visible in Extended Data Fig. 2. The first is that the quality of our model is limited at blue wavelengths by the signal-to-noise in the M8 template spectrum we have adopted for the pristine photosphere. Although our nuisance parameters can account for this to some degree, this is another reason why the statistical errors quoted on parameters are likely under-estimates. The second is that there are features in the residuals which are of similar amplitude to features in the difference spectrum. These features arise because the M8 template is not a perfect fit to the pristine photosphere. However, this does not mean that our fit to the difference spectrum is unreliable. If we label our adopted photosphere template P, and the true photosphere P', then the error in the high state spectrum will be

$$H - H' = N(1 - f_h)(P - P'),$$

whereas the error in the difference spectrum will be

$$\Delta - \Delta' = N(f_h - f_l)(P - P').$$

Thus, residuals in the high state spectrum will also be present in the difference spectrum, but reduced in size by a factor of more than 50; this implies the residuals will be smaller than the value we adopt for our nuisance parameter when fitting the difference spectrum.

**An orbiting exoplanet as an electrodynamic engine**

An orbiting planetary body embedded within the magnetosphere of LSR J1835 will have motion relative to this magnetosphere and any associated frozen-in plasma. If the planet is conducting, this motion leads to the generation of an electric field across the planet that can power auroral emissions on LSR J1835[28]. The expected power produced is proportional to the intercepted flux of magnetic energy, $P \propto vB_\perp^2 R_{obs}^2$, where $B_\perp$ is the component of the magnetic field perpendicular to the planet's orbital motion, $v$ is the planets velocity relative to the local magnetic field and $R_{obs}$ is the size of the obstacle created by the planet, the latter defined by its ionosphere or magnetosphere depending on whether the planet is magnetized or unmagnetized[17]. For example, a simple scaling from the Jupiter-Io system indicates that an unmagnetized Earth-sized planet orbiting within 20 radii (<30 hour orbital period) of LSR J1835 will generate a current sufficient to power the observed aurorae.

However, the resulting auroral emission is expected to be strongly modulated by the orbital period of the planet, whereas in the case of LSR J1835, the observed periodicity of 2.84 hours is consistent with rotation of the dwarf, as inferred from rotational broadening of spectral lines. Indeed, a planet orbiting with this period would be within the Roche limit of LSR J1835 and torn apart by tidal forces. An alternative possibility arises if the magnetic field at the location of the planet varies substantially as LSR J1835 rotates, which will be the case if LSR J1835 possesses a non-axisymmetric magnetic field. For example, a tilted dipole would result in as much as 4-fold variation in the current produced across the orbiting planet during each rotation of LSR J1835. The period of the auroral emission would then be $P_{aur} = P_{rot} (1 + P_{rot}/P_{orb})$ and would approach the rotation period for large values of $P_{orb}/P_{rot}$. In this scenario, the aurorae would display modulation close to the rotation period as well as on the orbital period.

**Code Availability**

The code used to model the auroral feature of LSR J1835 is publicly available at https://github.com/StuartLittlefair/lsr1835

31. Horne, K. Optimal Spectrum Extraction and Other CCD Reduction Techniques. *Publ. of the Astron. Soc. Pacific*, **98**, 609-617 (1986)

32. Bochanski, J. J., West, A. A., Hawley, S. L. & Covey, K. R. Low-Mass Dwarf Template Spectra from the Sloan Digital Sky Survey, *Astron. J.* **133,** 531-544 (2007)

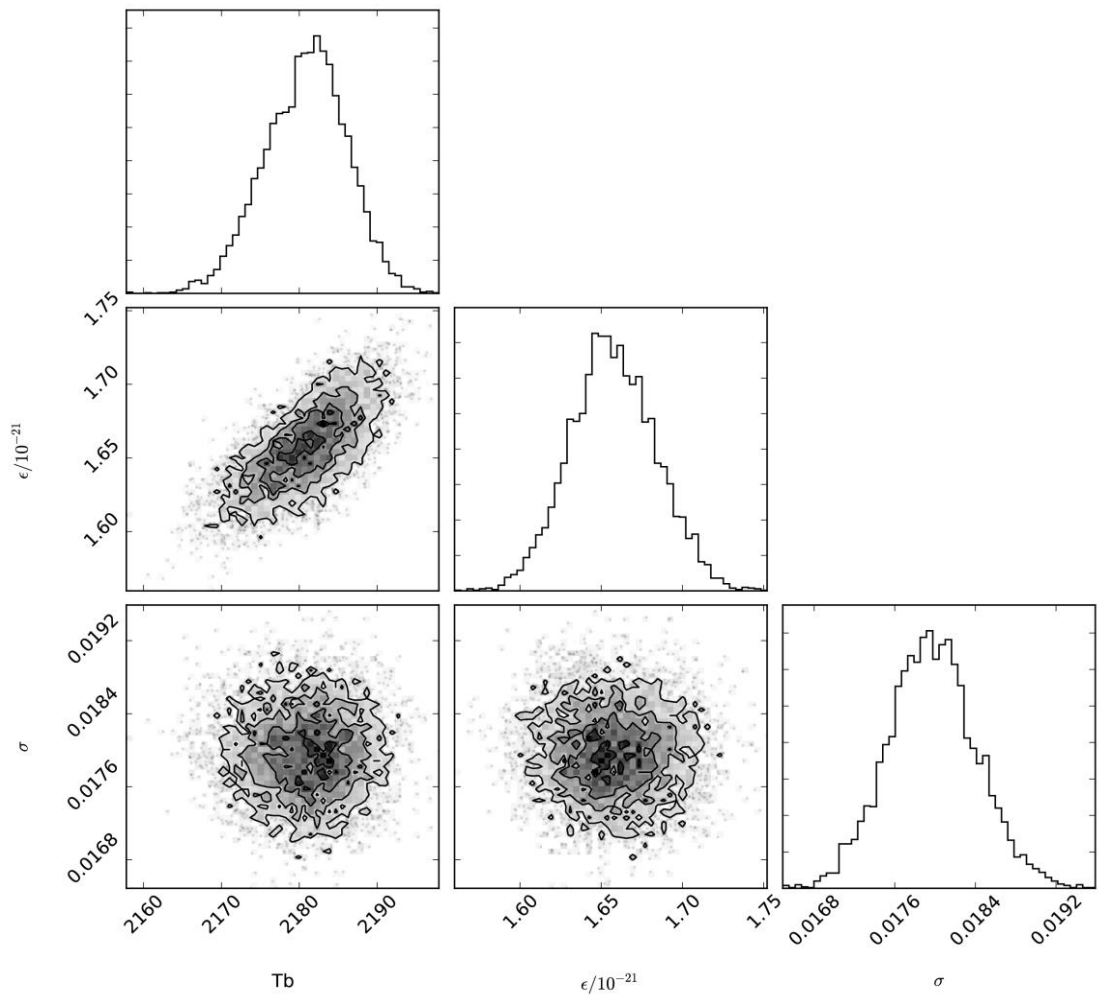

**Extended Data Figure 1 | Posterior probability distributions for two-phase model parameters.** Greyscales with contours show our estimates of the joint posterior probability distributions for all combinations of parameters, whilst marginal posterior distributions are shown with histograms.

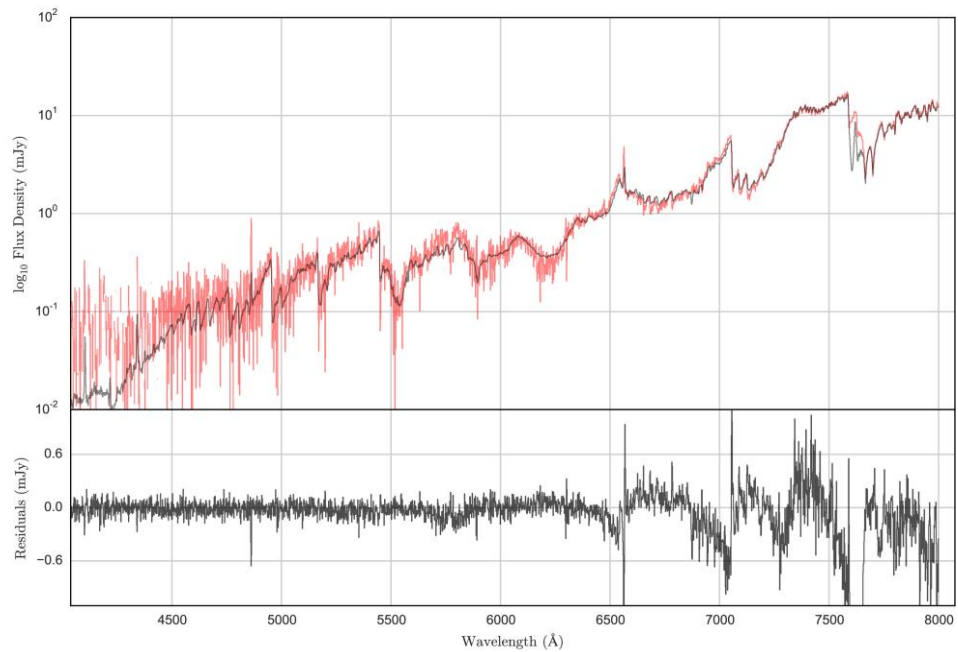

**Extended Data Figure 2 | The high state spectrum of LSR J1835. a.** The high state spectrum of LSR J1835 (black) is shown along with the best fit model to the high state (red). **b.** The residuals between the model and the fit.